# Insulator to Metal Transition, Spin-Phonon Coupling, and Potential Magnetic Transition Observed in Quantum Spin Liquid Candidate LiYbSe$_2$ under High Pressure

Haozhe Wang[1#], Lifen Shi[2,3#], Shuyuan Huyan[4,5], Greeshma C. Jose[6], Barbara Lavina[7,8], Sergey L. Bud'ko[4,5], Wenli Bi[6], Paul C. Canfield[4,5], Jinguang Cheng[2,3*], Weiwei Xie[1*]

1. Department of Chemistry, Michigan State University, East Lansing, MI 48824, USA
2. Beijing National Laboratory for Condensed Matter Physics and Institute of Physics, Chinese Academy of Sciences, Beijing 100190, China
3. School of Physical Sciences, University of Chinese Academy of Sciences, Beijing 100190, China
4. Ames National Laboratory, Iowa State University, Ames, IA 50011, USA
5. Department of Physics and Astronomy, Iowa State University, Ames, IA 50011, USA
6. Department of Physics, University of Alabama, Birmingham, AL 35294, USA
7. Center for Advanced Radiation Sources, The University of Chicago, Chicago, IL 60637, USA
8. Advanced Photon Source, Argonne National Laboratory, Argonne, IL 60439, USA

[#] H.W. and L.S. contributed equally. * Email: xieweiwe@msu.edu; jgcheng@iphy.ac.cn

*Abstract*

Metallization of quantum spin liquid (QSL) materials has long been considered as a potential route to achieve unconventional superconductivity. Here we report our endeavor in this direction by pressurizing a three-dimensional QSL candidate, LiYbSe$_2$, with a previously unreported pyrochlore structure. High-pressure X-ray diffraction and Raman studies up to 50 GPa reveal no appreciable changes of structural symmetry or distortion in this pressure range. This compound is so insulating that its resistance decreases below $10^5$ Ω only at pressures above 25 GPa in the corresponding temperature range accompanying the gradual reduction of band gap upon compression. Interestingly, an insulator-to-metal transition takes place in LiYbSe$_2$ at about 68 GPa and the metallic behavior remains up to 123.5 GPa, the highest pressure reached in the present study. A possible sign of magnetic or other phase transition was observed in LiYbSe$_2$. The insulator-to-metal transition in LiYbSe$_2$ under high pressure makes it an ideal system to study the pressure effects on QSL candidates of spin-1/2 Yb$^{3+}$ system in different lattice patterns.

# Introduction

The quantum spin liquid (QSL) state, a highly entangled superposition state of all the possible configurations of valence-bond spin singlets and the spin excitation continuum, was proposed to understand unconventional superconductivity in high $T_c$ superconductors.[1-4] The geometrically frustrated spin-1/2 ($S = 1/2$) systems, such as two-dimensional Kagome, triangular, and three dimensional pyrochlore lattices, are ideal structural platforms to host QSL states.[5-8] The rare-earth-based frustrated magnetic systems, such as YbMgGaO$_4$ with $S = 1/2$ Yb$^{3+}$ triangular lattice, were proposed and studied recently as promising QSL candidates.[9-11] However, the intrinsic structural disorder with a random distribution of Mg$^{2+}$ and Ga$^{3+}$ on one atomic site in YbMgGaO$_4$ may generate a spin-liquid-like state at low temperatures.[12] Thus, NaYbCh$_2$ (Ch = O, S, and Se), which contain a perfect Yb$^{3+}$ triangular lattice, have been proposed and widely explored as an $S = 1/2$ rare-earth-based system without inherent atomic disorders in crystals.[13-15] Our recent work reported that LiYbSe$_2$, an isoelectronic analogue of NaYbSe$_2$, suggests a QSL state in which Yb$^{3+}$ adopts an effective $S = 1/2$ pyrochlore lattice.[16]

High pressure technique has been used as a clean tool to induce unconventional superconductivity in QSL candidates.[17,18] For example, an approximate triangular lattice of $S = 1/2$ Cu$^{2+}$ ions in the organic compound $\kappa$-(BEDT-TTF)$_2$Cu$_2$(CN)$_3$ was proposed theoretically and experimentally examined to host a QSL state.[19] The electron spin resonance (ESR) measurements down to 50 mK indicated an absence of magnetic ordering.[20] By applying a very small pressure (~0.35 GPa), the insulator-to-metal transition was observed at around 13 K and followed by a superconducting transition at around 4 K.[21,22] Recently, high pressure study was reported on another QSL candidate, NaYbSe$_2$.[23,24] A structural study using X-ray diffraction and Raman scattering up to 30 GPa indicates a subtle and collapse-like structural change or valent change on Yb$^{3+}$ without symmetry change at around 12 GPa. The electrical resistance is gradually reduced by applying pressure with a superconducting phase transition occurring at ~5.8 K and ~20 GPa. The metallization trend is confirmed by further high-pressure measurements. These studies inspired us to investigate the resistance changes and possible superconductivity under high pressure in other QSL candidates.[23]

Herein, we present the high-pressure studies including synchrotron X-ray diffraction (XRD), Raman spectroscopy, electrical resistance measurements, and electronic band structure calculations on LiYbSe$_2$. There were no structural changes in LiYbSe$_2$ observed up to 50 GPa. Interestingly, the resistance measurements confirmed the insulator-to-metal transition occurs around 68.0 GPa, and the occurrence of possibly magnetic phase transition was detected above 90 GPa.

## Experimental Details

**Crystal Growth:** The synthesis of LiYbSe$_2$ was carried out through the LiCl salt flux growth method, which is a widely employed technique for the synthesis of single crystals. The elemental constituents, namely Yb powder (99.9%, Alfa Aesar), Se shots (99.999%, BTC), and the flux LiCl (98+%, Alfa Aesar), were mixed in the molar ratio of 1:2.2:20. The resulting mixture was placed in a crucible made of silica, covered with quartz wool, and sealed inside an evacuated tube. The tube was heated to 400 °C and annealed for 3 hours, then the temperature was slowly ramped up to 850°C and the tube annealed for one week before cooling down to room temperature at a rate of 15°C/hr. Subsequently, the sample was washed with distilled water to remove any excess flux and air-dried before further measurements were performed. Single crystals of LiYbSe$_2$ were used for the high pressure study.

**Synchrotron X-ray Diffraction (XRD) under high pressure:** The synchrotron powder XRD experiments were carried out up to 15.3 GPa at the Beamline 3ID of the Advanced Photon Source, Argonne National Laboratory. X-rays with a wavelength of 0.4833 Å were focused to 15 μm size. Powder sample ground from single crystals was loaded in a BX90 DAC with anvils of 400 μm diameter culet size.[25] Neon was loaded as pressure transmitting medium (PTM) and Ruby was used to measure pressure in-situ.[26] The 2-D diffraction images were integrated using DIOPTAS software[27] and Rietveld refinements on the XRD data were performed in GSAS-II[28].

**High-pressure Raman Spectrum:** To monitor the crystal phase information and bonding interactions in LiYbSe$_2$, the Raman spectra were collected by using a Raman system (Spectroscopy & Imaging) with 633 nm laser excitation. The 4:1 methanol-ethanol was used as the PTM and the pressure in DAC was monitored by the R1 fluorescence line of ruby up to 48.3 GPa.[29,30]

**High Pressure Electrical Resistance Measurements:**
**Ambient pressure to 58.2 GPa pressure range measured at Ames National Laboratory:** Electrical resistivity measurement by Van der Pauw method was performed in commercial Diamond Anvil cell that fits Quantum Design Physical Property Measurement System (PPMS). 300 μm culet-size standard-cut diamonds were used as anvils. LiYbSe$_2$ single crystal of dimension close to 45 × 45 μm$^2$ was directly picked out from the sample batch and polished into thin flakes with a thickness of about 15 μm. The sample was loaded together with a tiny piece of ruby ball into the 250 μm thick, apertured, stainless-steel gasket, which was pre-indented to ~35 μm, covered by *c*-BN. The sample chamber was about 120 μm in diameter. Platinum foil was used as the electrodes to connect the sample. Mineral oil was used as PTM to provide a better hydrostatic environment at low pressure range. Pressure was determined by the R1 line of the ruby florescent spectra. Low temperature resistance measurements down to 1.8 K were conducted in the Quantum Design PPMS.

**68.0 GPa to 123.5 GPa pressure range measured at Institute of Physics, CAS:** To measure resistance to higher pressure, diamond anvil cells manufactured by HMD with anvils of smaller culet sizes of 200 μm and 100 μm were used. In the first run from 68.0 to 94.2 GPa, the rhenium gasket was pre-indented to ~34 μm and then a 60 μm-diameter hole was drilled in the center using a laser drilling system. The rhenium gasket was covered with a *c*-BN epoxy insulating layer. A piece of LiYbSe$_2$ single crystal with dimensions of about $58 \times 29 \times 12$ μm$^3$ was placed at the center of the sample chamber filled with a KBr PTM.[31] In the second run from 68.4 to 123.5 GPa, the sample with a size of $29 \times 29 \times 5$ μm$^3$ was placed directly on the *c*-BN epoxy insulating layer at the center of the pre-indented gasket culet without PTM used at all. The pressure was determined by the Raman spectrum of diamond in the whole pressure range.

**Electronic Structure Calculation:** The Vienna Ab initio Simulation Package (VASP) version 5.4.4, which implements the projector augmented wave (PAW) method, was used to generate the electronic band structure, density of states (DOS), and Fermi surface.[32,33] The structure was taken from the high-pressure powder XRD refinement. The generalized gradient approximation (GGA) parameterized Perdew-Burke-Ernzerhof (PBE) functional was used to account for the electronic exchange and correlation.[34,35] Spin polarization using GGA+$U$ ($U_{\text{eff}}$ = 6 eV)[36] was employed to treat Yb *f* electrons as valence electrons. A kinetic energy cutoff of 800 eV was used for the plane-wave basis set. Γ-point centered Monkhorst–Pack *k*-point grids of $8 \times 8 \times 8$, $12 \times 12 \times 12$, $16 \times 16 \times 16$ were applied to sample the Brillouin zone for self-consistent field calculation, DOS, and Fermi surface, respectively. The energy converged to $10^{-8}$ eV. Spin-orbit coupling (SOC) was included in the calculation once specified. VASPKIT package was used for data processing.[37]

## Results and Discussions

**Robust Crystal Structure at High Pressure:** The crystal structural evolution of LiYbSe$_2$ under hydrostatic pressure was studied by performing XRD experiments at room temperature up to 15 GPa. As shown in **Fig. 1a**, the cubic pyrochlore structure with the space group *Fd*-3*m* is maintained under pressures up to 15 GPa. The volume of the unit cell obtained from the Rietveld refinements is plotted as the function of pressure in **Fig. 1b**, which shows a smooth contraction upon compression with the lattice parameter *a* changing from 11.242(1) Å at ambient pressure to 10.656(1) Å at 15.3 GPa, further confirming the absence of any structural transition in the studied pressure range. A least-square fitting to the volume vs pressure data using the second-order Birch-Murnaghan equation of state (EOS), **Eqn. 1**, yields the isothermal bulk modulus $K_{T_0}$ = 61(2) GPa, and the volume at zero pressure $V_0$ = 1445(5) Å$^3$. The third-order Birch-Murnaghan EOS, **Eqn. 2**, was also applied to give the bulk modulus $B_0$ = 83(6) GPa, the pressure derivative $B_0'$ = 1.7(6) and the volume at zero pressure $V_0$ = 1424(5) Å$^3$.[38] It should be noted that the standard error of the volumes here is of ~0.4 Å$^3$ and not visible in the **Fig. 1b**.

$$P(V) = \frac{3}{2} K_{T_0} \left[ \left(\frac{V_0}{V}\right)^{\frac{7}{3}} - \left(\frac{V_0}{V}\right)^{\frac{5}{3}} \right] \quad (1)$$

$$P(V) = \frac{3}{2} B_0 \left[ \left(\frac{V_0}{V}\right)^{\frac{7}{3}} - \left(\frac{V_0}{V}\right)^{\frac{5}{3}} \right] \left\{ 1 + \frac{3}{4}(B_0' - 4) \left[ \left(\frac{V_0}{V}\right)^{\frac{2}{3}} - 1 \right] \right\} \quad (2)$$

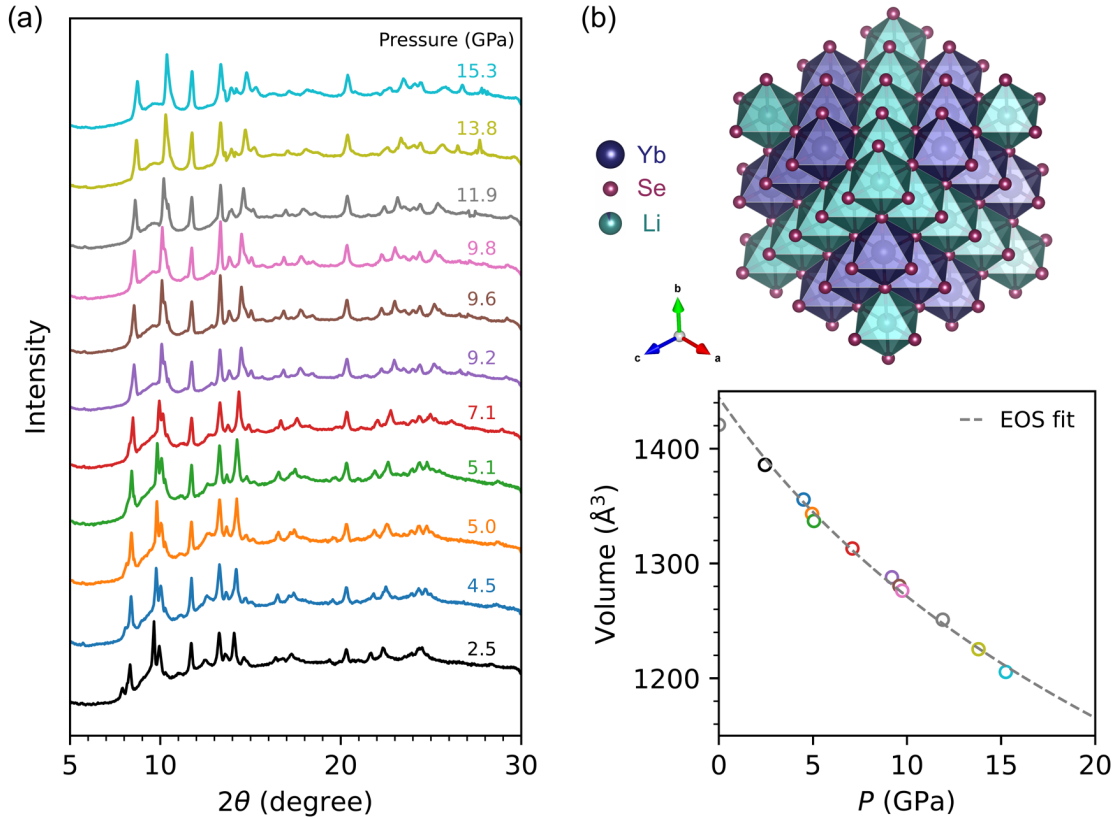

**Fig. 1** (*a*) High pressure XRD data taken at room temperature up to 15.3 GPa. (*b*) Pressure dependence and second-order Birch-Murnaghan EOS fit of the unit cell volume of LiYbSe$_2$. The crystal structure is also shown.

**Crystalline Electric-Field Excitation and Spin-Phonon Coupling Observed from Raman Spectra:** To investigate the bonding vibration under high pressure, the Raman spectra under high pressure were taken and summarized in **Fig. 2**. There are three Raman active modes observed at ~235 (Mode 1), ~175 (Mode 2), and ~128 (Mode 3) cm$^{-1}$ at ambient pressure, which is consistent with the Raman active modes of a cubic *Fd*-3*m* structure (origin choice 2) with the 16*c*, 16*d*, and 32*e* sites occupied (A$_{1g}$, E$_g$, T$_{2g}$). Modes 2 and 3 smoothly move to higher frequencies as the pressure increases. The changes in Raman modes are summarized in **Fig. 2b**. At ~5.6 GPa, a small peak at 207 cm$^{-1}$ appears. It moves to lower frequency under pressurization, merging with Mode 2 at about 200 cm$^{-1}$ and 8.6 GPa. According to the high-pressure X-ray diffraction, no structural transition was observed up to 15.3 GPa, thus the appeared new mode is not from the structural distortion. At about 11.2 GPa, it becomes significantly observable, which may also be considered as the continuous excitation peak from Mode 1 crossing Mode 2 since there is no structural transition around that pressure. The intensity of this peak enhances at a higher pressure and moves to lower frequencies until merged with the peaks of Mode 3 at around ~180 cm$^{-1}$ and 19.3 GPa. On the other hand, the very weak peaks at high pressure range can be considered as the extension of Mode 1 shown in **Fig. S1**. The phonon coupling has been reported in many rare-earth compounds in pyrochlore structure oxides.$^{39}$ In LiYbSe$_2$ under high pressure, the anomalies Raman peak may be caused by the CEF-phonon coupling, also named spin-phonon coupling instead of being the continuous peak from Mode 1. The interaction between spins and phonons, commonly known as spin-phonon coupling, can be attributed to the presence of a broad and strong Mode 2, which is considerably larger in magnitude when compared to the exchange coupling. The broadening of this mode is attributed to resonance and overlaps with phonon frequencies, particularly under high pressure conditions.$^{39}$ As the pressure increases, the peaks in Raman spectra become broader and weaker as the crystallinity decreased, without the emergence of any additional peaks up to 48.3 GPa. As shown in **Fig. 2c**, the crystal looks transparent red and turns darker and darker with pressure applied. Once the system was back to ambient pressure, there was only one remaining Raman peak, indicating decreased crystallinity. This could also explain the irreversible color change.

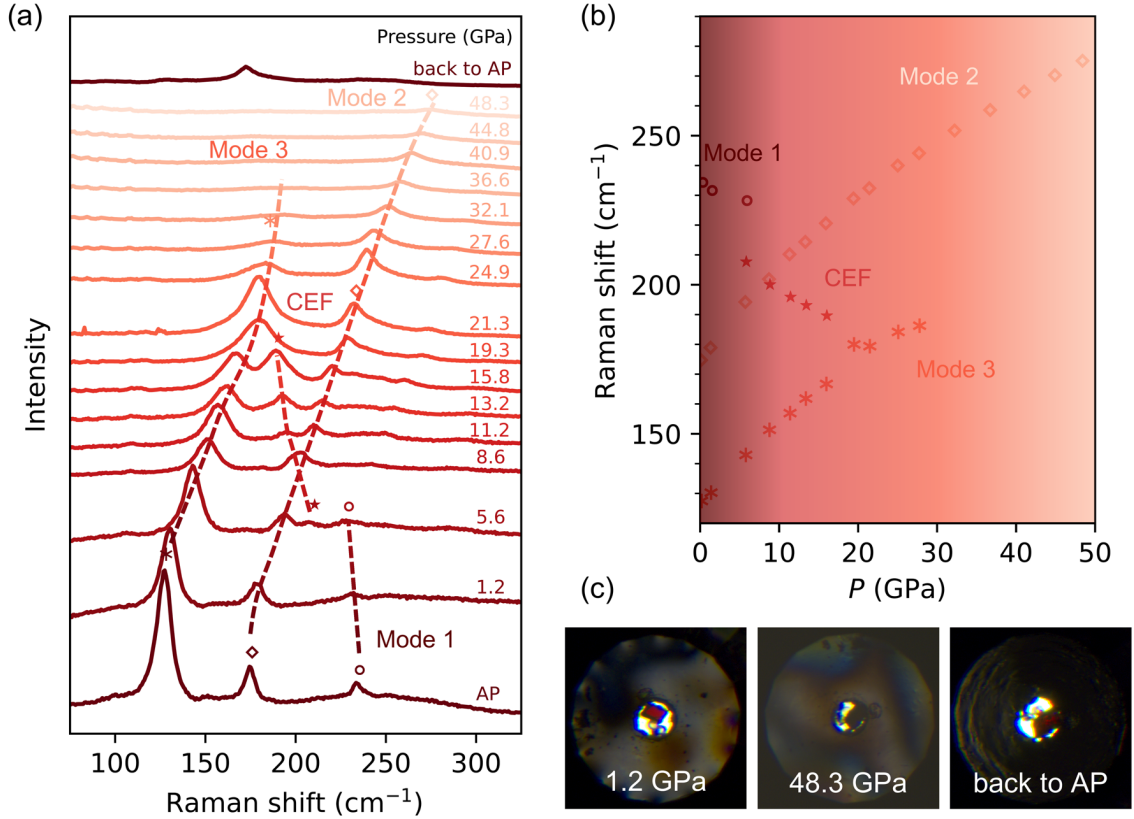

**Fig. 2 (*a*)** High pressure Raman spectra of LiYbSe$_2$ up to 48 GPa. (***b***) Evolution of Raman vibration modes under pressure. (***c***) Optical image of the sample under inside the DAC. The color of the crystal becomes darker when increasing the pressure and goes back once the pressure decreases.

**Insulator to Semiconductor to Metal Transitions:** To study the electrical transport properties and explore potential superconductivity in LiYbSe$_2$, the temperature dependence of resistance *R* of LiYbSe$_2$ between 2 and 300 K under various pressures up to 58.2 GPa was measured. The results are shown in **Fig. 3*a***. At lower pressure range (< 22 GPa), the resistance is too large to be measured in the PPMS. Starting from 22.2 GPa, the resistance can be detected. From 22.2 to 58.2 GPa, LiYbSe$_2$ shows semiconductor-like behavior in the whole temperature range. The resistivity can be described with the Arrhenius activation model, expressed as **Eqn. 1**, in which *A* is a parameter related to the system, $\Delta_g$ is the semiconductor band gap, *T* is temperature, and $k_B$ is Boltzmann constant.

$$\rho = A e^{\frac{\Delta_g}{2k_B T}} \tag{1}$$

To evaluate the band gap changes under the pressure, the resistance in the temperature range of 165 K to 250 K was fitted accordingly. As shown in **Fig. 3b**, the resistance of the system is roughly described by the activation model, presented the near-linear correlation between the logarithm of resistance and $10^3/T$ within the range of temperatures under examination. The band gaps $\Delta_\rho$, under different pressures, inferred from these data are presented in **Fig. 3c**. In general, the band gap monotonically decreases with pressure; however, a sudden drop was observed at around 30 GPa, where it decreased from 0.26 eV to below 0.10 eV. This could also be indicated by a dramatic slope change in **Fig. 3b** between 28.4 GPa and 31.6 GPa ones. At ambient pressure, the Yb in LiYbSe$_2$ shows 3+ oxidation state, together with the disappearance of Raman active Mode 3 in **Fig. 2**, it may indicate that there is a subtle phase transition such as ~~valent changes in Yb~~ magnetic transition in the system at this pressure. As pressure increases further, the band gap decreases slightly to ~0.02 eV at 58.2 GPa.

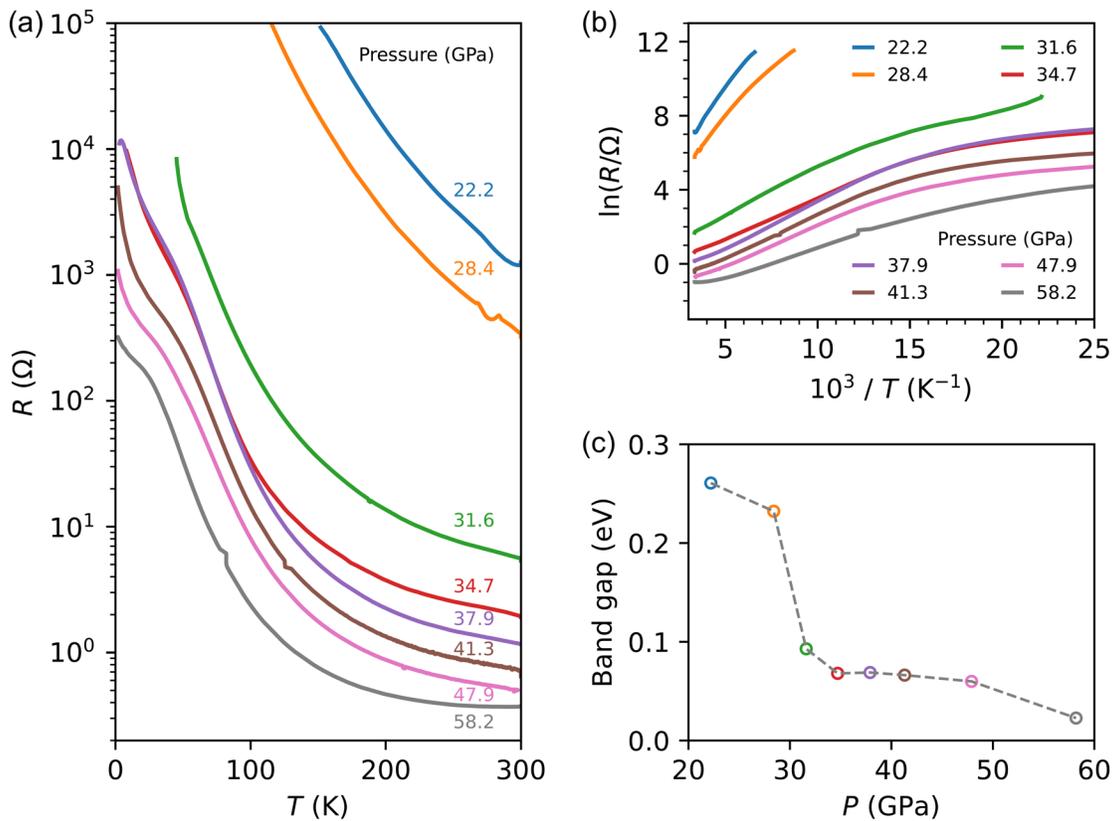

**Fig. 3** (*a*) Temperature dependence of resistance of LiYbSe$_2$ under different pressures. The semiconducting behavior was observed between 22.2 GPa and 58.2 GPa. (*b*) The logarithm of resistance plotted as a function of $10^3/T$. (*c*) Pressure dependence of band gap derived from the fit in (*b*).

To investigate the electrical resistance changes under higher pressure, further studies were conducted, as shown in **Fig. 4**. At 68.0 GPa, the metallic behavior was observed with a minimum of around 25 K, which may be caused by weak localization effect originating from the presence of disorder potentials and incoherent Kondo effect due to the presence of $Yb^{3+}$ mixture on $Li^+$ site.[16,23] As the pressure increases to 80 GPa, the minimum disappears and a complete metallic behavior in the whole temperature range (1.8–300 K) is achieved. When the experiments were repeated without the pressure medium, the minimum in resistivity was found to be absent. Instead, only a transition from an insulator to a metallic state was observed, as depicted in **Fig. S2**. Such progressive changes at low temperatures and high pressures were also observed in other Yb-series heavy-fermion compounds.[40] $LiYbSe_2$ shows metallic behavior up to 94.2 GPa. The metallization under pressure is also confirmed by the crystal color changes under pressure in **Fig. 4**. Similar to **Fig. 2c**, the crystal color starts from transparent red and changes to darker and darker with pressure, indicating the gradual increase of metallicity. Surprisingly, at 94.2 GPa, a very weak but sudden resistance drop was observed around 5.0 K; in addition, this drop can be eliminated by the application of 0.6 T magnetic field. Given the very small size of this drop, the most likely explanation is a tiny amount of second phase (possibly Se) that becomes superconducting. Of course, a more enticing option would be the onset of a bulk transition that needed high temperatures to complete. These results motivated us to conduct higher-pressure measurements to explore the potential superconducting or magnetic transition that exists in $LiYbSe_2$.

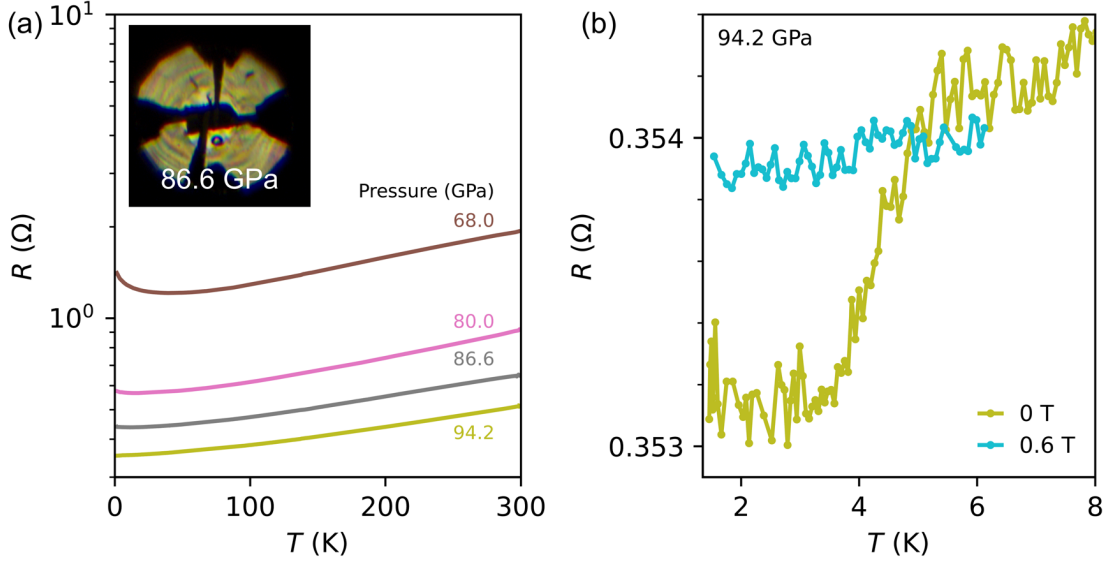

**Fig. 4** (*a*) Temperature dependence of resistance of LiYbSe$_2$ under different pressure. The system becomes metallic when pressure of 68.0 GPa is applied. (*b*) Temperature dependence of the resistance at 94.2 GPa with 0.6 T magnetic field applied.

**Potential Phase Transition Observed above 96.0 GPa:** The resistance was measured again in another run without PTM, having the pressure applied up from 68.4 GPa to 123.5 GPa. The results are presented in **Fig. 5a**. As the pressure increased, a very weak drop in resistance was observed starting at around 3.0 K from 96 GPa, shown in **Fig. 5b**. The potential phase transition temperature is slightly suppressed with pressure from 3.0 K at 96 GPa to around 2.5 K at 116.7 GPa. Nevertheless, the size of the anomaly is insufficient to draw any definitive conclusion. At 123.5 GPa, the field-dependent resistance shows that the resistance drop is gradually suppressed. At 0.1 T, the transition is completely suppressed, indicating the drop on resistance curve at around 3.0 K may originate from the magnetic (antiferromagnetic or some short-range magnetic order) transition. These results strongly suggest that the feature we see is, at best, a very small second phase that might have a superconducting transition and does not represent a bulk phase. It has been observed that the datasets exhibit discrepancies in the overlapping pressure range, shown in **Fig. S4**, which can be attributed to variations in the levels of non-hydrostaticity between the two experiment runs. Despite this observation, it is noteworthy that these deviations do not compromise the validity or integrity of our overall conclusion.

In order to understand the metallic state and potential phase transition under pressure, we fitted the resistance data with power-law equation $R(T) = R_0 + AT^n$, where $R_0$ is the residual resistance, and the pre-factor $A$ is a parameter related to electron-electron interactions and electron-phonon coupling. **Fig. 5d** shows that the power-law formula fits the experimental data well in the temperature range between 10 K and 100 K. The pressure dependence of the exponent $n$ is plotted in **Fig. 5e**. At the boundary of insulator-to-metal transition, the metallic state of LiYbSe$_2$ apparently shows a Fermi-liquid (FL) ground state behavior with $n \sim 2$. With pressure increasing, the $n$ decreases gradually and approaches $n = 1.68(4)$ at $P = 123.5$ GPa. Therefore, there might be a crossover from FL to non-Fermi liquid (NFL) behavior with the increase in pressure as approaching the boundary of possible superconductivity or magnetic transition.

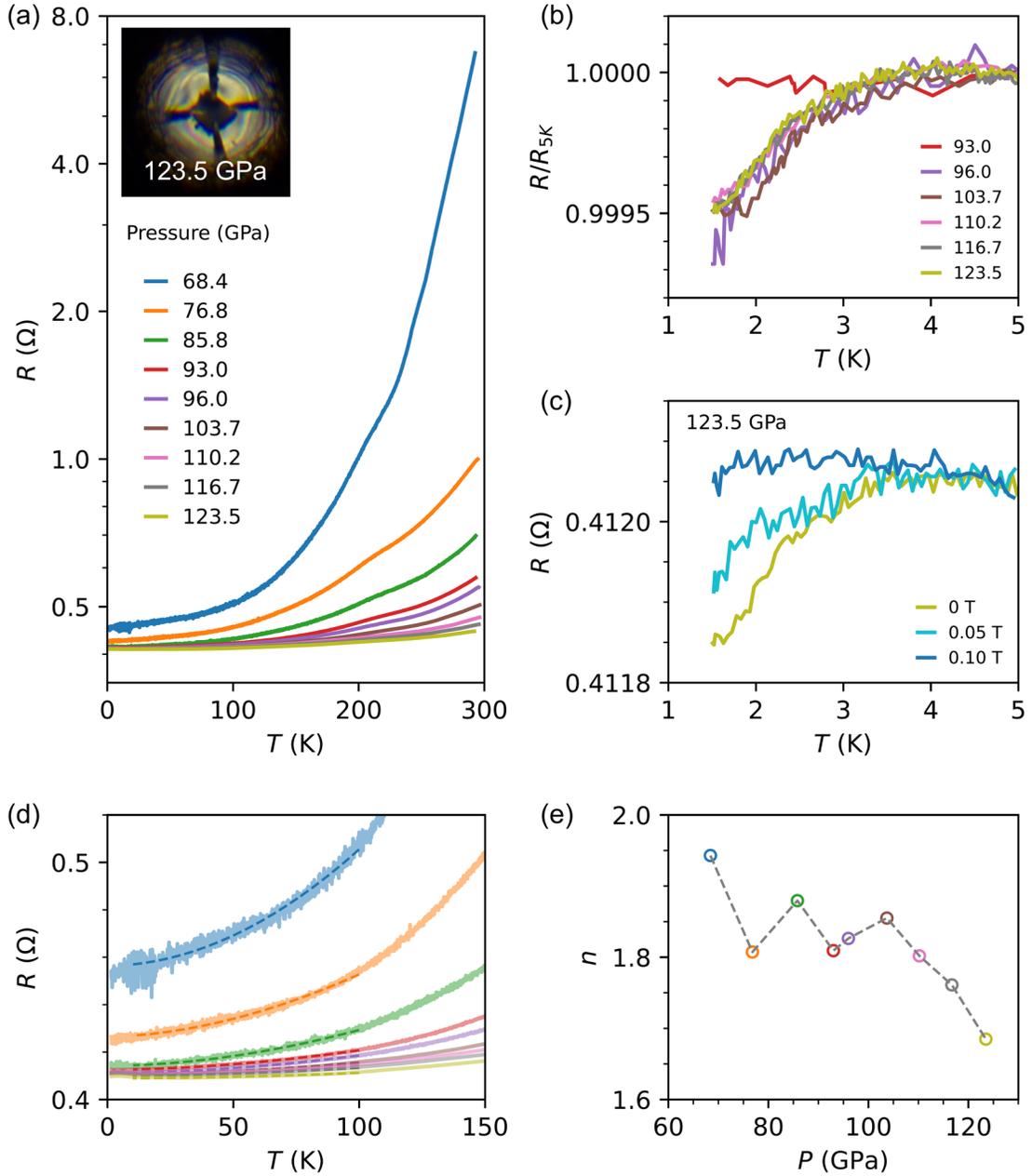

**Fig. 5** (*a*) Temperature dependence of resistance of LiYbSe$_2$ under pressure up to 123.5 GPa. Insert, optical photo of the DAC in the experiment under 123.5 GPa. (*b*) Normalized resistance at low temperatures under pressure. (*c*) Field dependence of the resistance at 123.5 GPa. (*d*) Power-law fitting of the resistance under different pressures in the range of 10 K to 100 K. (*e*) Pressure dependence of fitted $n$ in metallic state using the power-law equation.

## Conclusion

The high-pressure studies on LiYbSe$_2$ in cubic pyrochlore lattice are presented. As pressure is applied, the spin-phonon interactions occur in LiYbSe$_2$ at around 6 GPa according to the Raman spectra. High pressure powder X-ray diffraction confirms the peak appearing in Raman spectra originates from the crystal electric field excitation rather than from the structural distortion. The resistance measurements show the insulator-to-metal transition around 68 GPa. A low temperature field-dependent anomaly was detected in the resistance measurements at and above 94.2 GPa, which may be related to magnetic or structural transition. However, under pressure up to 123 GPa, no complete structural, or even magnetic transition was observed, and we conclude that this feature is most likely associated with a small amount of second phase rather than representative of a bulk property. Apparently, the subtle differences of Yb$^{3+}$ lattices in LiYbSe$_2$ play a critical role in resulting in significantly different physical properties from other quantum spin liquid candidates. LiYbSe$_2$ can provide an ideal platform to study the structure-property relationship in quantum spin liquid candidates.

## Acknowledgment

The work at Xie's group was supported by U.S. DOE-BES under Contract DE-SC0023648. H.W. and W.X. thank the computing resource support from Prof. Gabriel Kotliar at Rutgers University Parallel Computing (RUPC), Center of Materials Theory, Department of Physics and Astronomy. J.-G. C. is supported by the NSFC, MOST, and CAS through projects (Grant Nos. 12025408, 11921004, 2018YFA0305700, XDB25000000). Part of the high-pressure experiments used the Synergic Extreme Condition User Facility (SECUF). Work at the Ames National Laboratory (S.H., S.L.B., and P.C.C.) was supported by the U.S. Department of Energy, Office of Science, Basic Energy Sciences, Materials Sciences and Engineering Division under Contract No. DE-AC02-07CH11358. G. J. and W. B. acknowledge the support from the National Science Foundation (NSF) CAREER Award No. DMR-2045760. This research used resources of the Advanced Photon Source a U.S. Department of Energy (DOE) Office of Science user facility operated for the DOE Office of Science by Argonne National Laboratory under Contract No. DE-AC02-06CH11357. Use of the COMPRES-GSECARS gas loading system was supported by COMPRES under NSF Cooperative Agreement EAR -1606856 and by GSECARS through NSF grant EAR-1634415 and DOE grant DE-FG02-94ER14466.

# Supporting Information

# Insulator to Metal Transition, Spin-Phonon Coupling, and Potential Magnetic Transition Observed in Quantum Spin Liquid Candidate LiYbSe$_2$ under High Pressure

Haozhe Wang[1#], Lifen Shi[2,3#], Shuyuan Huyan[4,5], Greeshma C. Jose[6], Barbara Lavina[7,8], Sergey L. Bud'ko[4,5], Wenli Bi[6], Paul C. Canfield[4,5], Jinguang Cheng[2,3*], Weiwei Xie[1*]

1. Department of Chemistry, Michigan State University, East Lansing, MI 48824, USA
2. Beijing National Laboratory for Condensed Matter Physics and Institute of Physics, Chinese Academy of Sciences, Beijing 100190, China
3. School of Physical Sciences, University of Chinese Academy of Sciences, Beijing 100190, China
4. Ames National Laboratory, Iowa State University, Ames, IA 50011, USA
5. Department of Physics and Astronomy, Iowa State University, Ames, IA 50011, USA
6. Department of Physics, University of Alabama, Birmingham, AL 35294, USA
7. Center for Advanced Radiation Sources, The University of Chicago, Chicago, IL 60637, USA
8. Advanced Photon Source, Argonne National Laboratory, Argonne, IL 60439, USA

[#] H.W. and L.S. contributed equally. * Email: xieweiwe@msu.edu; jgcheng@iphy.ac.cn

## Table of Contents

**Fig. S1** High pressure Raman spectra of LiYbSe$_2$ up to 48 GPa. (*: mode 3; ◊: mode 2; ○: mode 1; star: crystalline electric-field excitation (CEF))

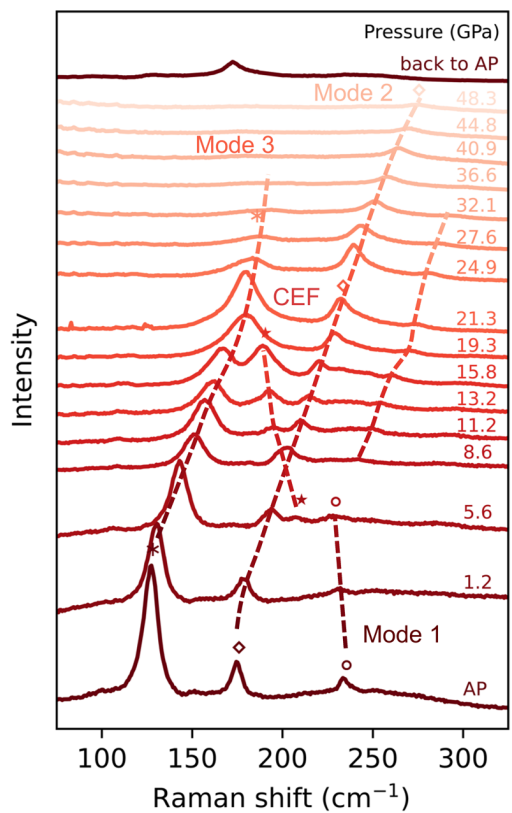

**Fig. S2** Temperature dependence of resistance of LiYbSe$_2$ under pressure. The data in the semiconducting region measured at Ames National Laboratory and Institute of Physics, CAS were highlighted in blue and red, respectively.

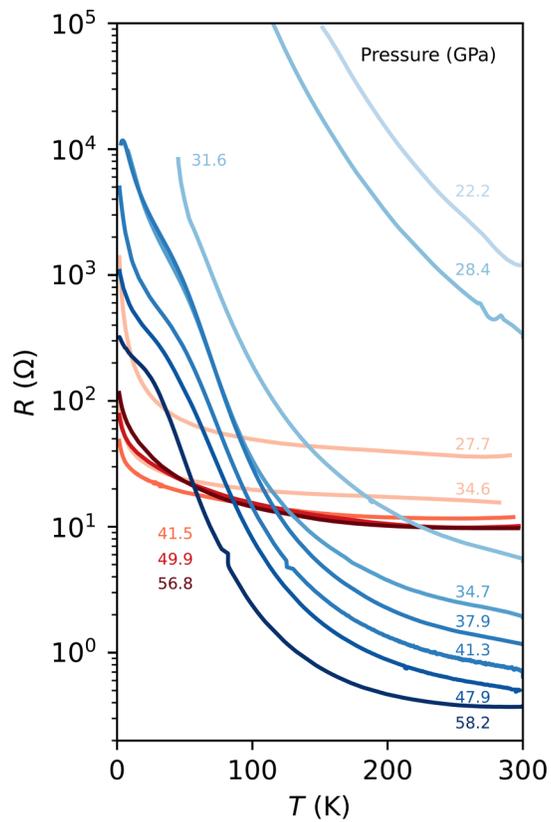

**Fig. S3** (*a*) Temperature dependence of resistance of LiYbSe$_2$ under different pressures. The semiconducting behavior was observed between 22.2 GPa and 58.2 GPa. (*b*) The logarithm of resistance in the temperature range of 160 K to 250 K plotted as a function of $10^3/T$. (*c*) Pressure dependence of band gap derived from the fit in (*b*).

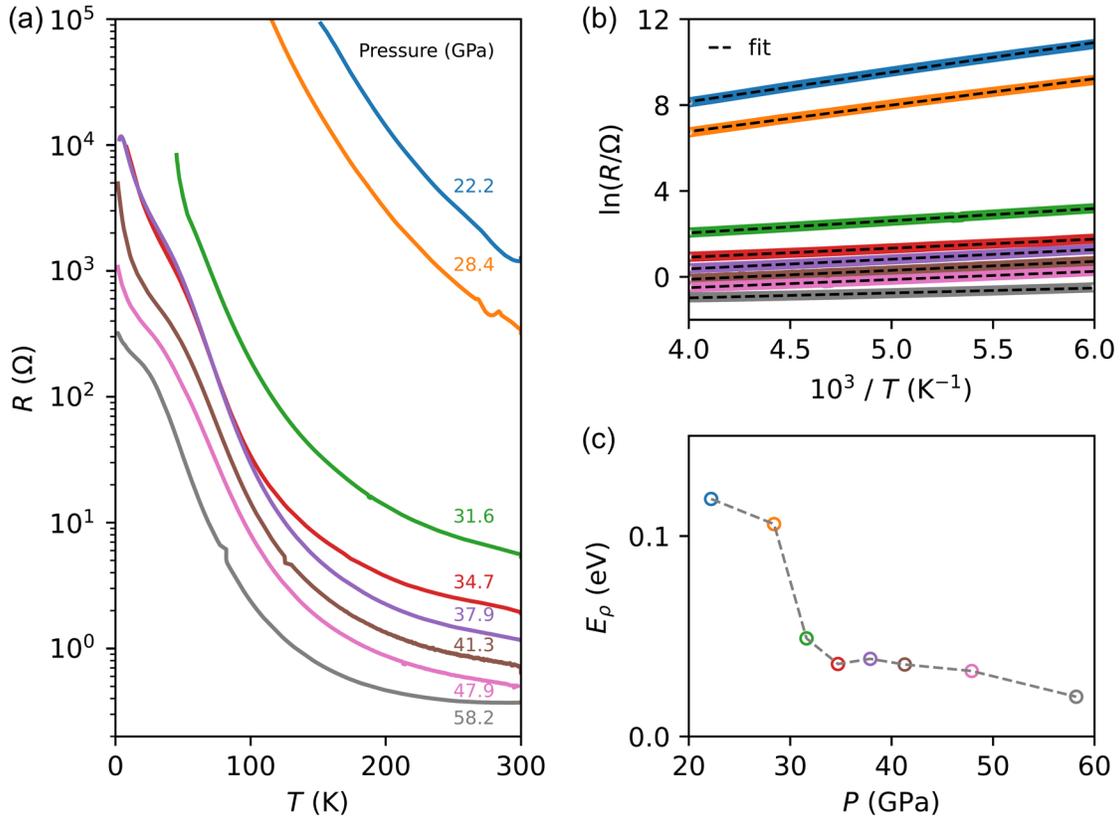

**Fig. S4** Temperature dependence of resistance of LiYbSe$_2$ under pressure. The data in the semiconducting region measured at Institute of Physics, CAS with/without pressure medium.

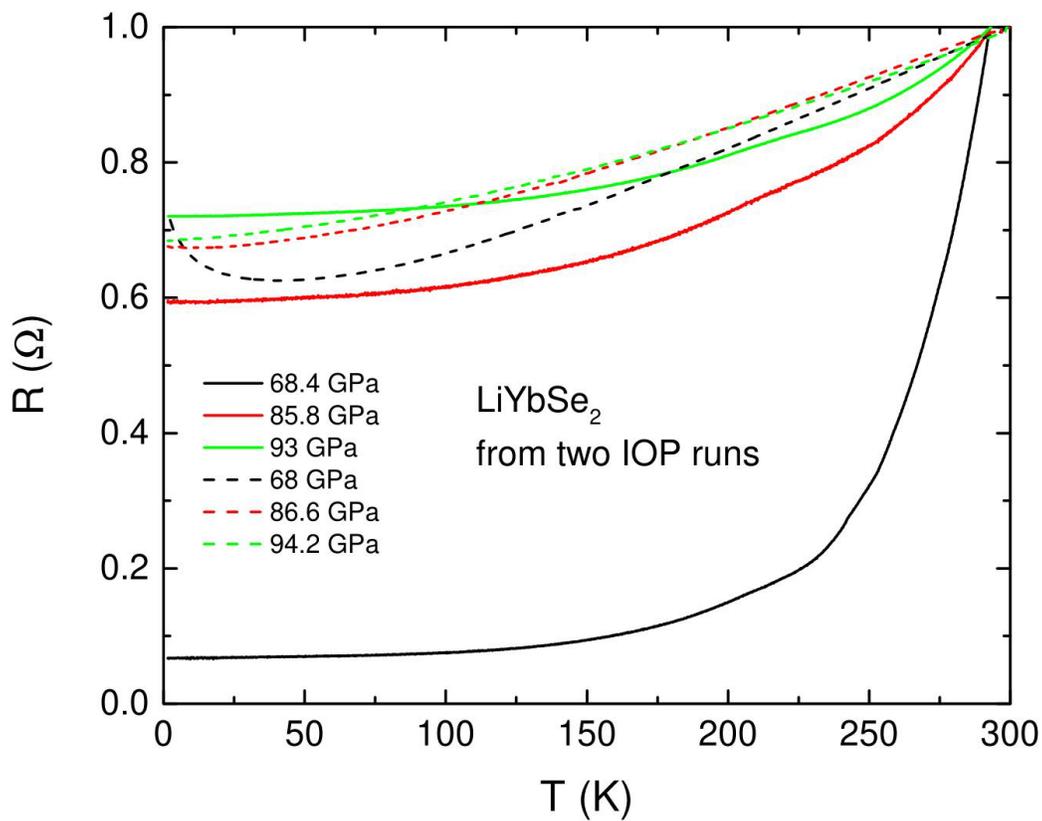

**Electronic Structures of LiYbSe$_2$ at 15 GPa:** To further understand the electronic structure of high pressure LiYbSe$_2$, spin polarization calculations using GGA, GGA+SOC, GGA+$U$ ($U_{eff}$ = 6 eV), and GGA+SOC+$U$ method were performed at 15 GPa. Here the effective Hubbard $U_{eff}$ is defined as $U_{eff} = U - J$, which incorporates the exchange interaction $J$ to the Coulomb term $U$.[1] This parameter is system-dependent and the accuracy of the calculation hinges on the choice of $U_{eff}$. The lattice parameters were taken from the Rietveld refinements of high pressure XRD. **Fig. 6** shows the calculated band structure. Based on $U$ = 6 eV, we find that Yb 4$f$ bands are located around -1.0 to -2.0 eV below the Fermi level. Besides, the Yb $d$, $f$ orbitals, and Se $p$ orbitals seem to hybridize, which contributes to the bands at around the Fermi level, with almost no Li contribution observed. Moreover, with the SOC effect included, the Yb 4$f$ bands were split, which interact with the Se bands near the Fermi level to generate a flat band at $\Gamma$ at around 0.2 eV below the Fermi level. Flat bands that are observed in the geometrically frustrated lattices, such as in twisted bilayer graphene, Kagome, and pyrochlore lattices, can host correlated electronic states, for example, ferromagnetism and superconductivity.[2,3] However, no saddle point was observed in the band structure which may explain the absence of superconductivity in the system. We have to mention that there is an inconsistency between experimental data and calculations here. At 15 GPa, our experimental data shows semiconducting behavior while the calculations give metallic system. LiYbSe$_2$ shows a possible QSL behavior at low temperature, which makes it difficult for a converged magnetic state to describe the system and the inconsistency may come from this. With the intrinsic complex electronic structure of LiYbSe$_2$, further theoretical studies may help to understand the system better.

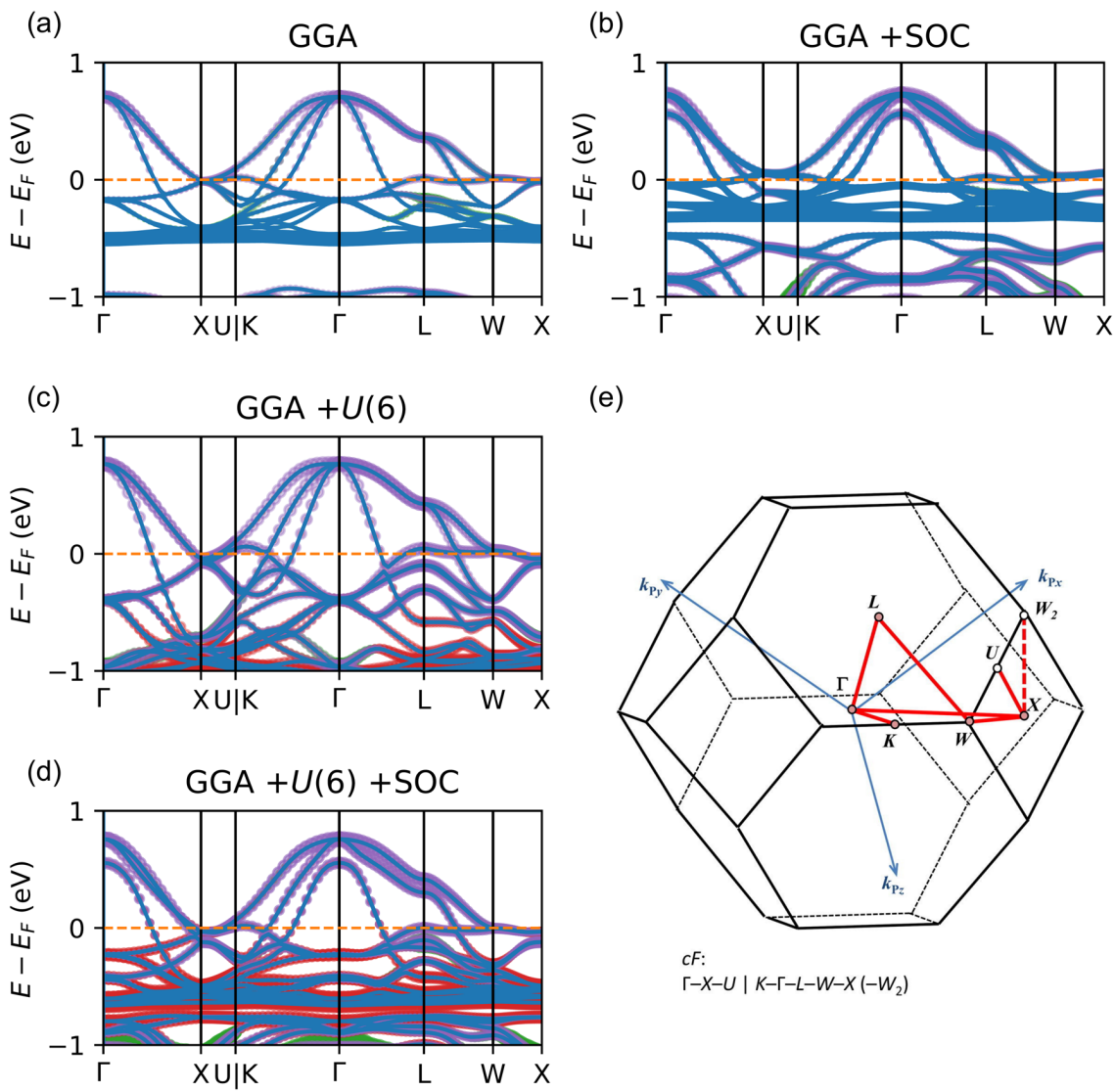

**Fig. S16** Calculated band structures of high pressure LiYbSe$_2$ with (**a**) GGA, (**b**) GGA+SOC, (**c**) GGA+$U$ ($U$ = 6 eV), and (**d**) GGA+$U$+SOC method. The Fermi level was highlighted by the orange dash line. Yb $d$ (green), $f$ (red) and Se $p$ (purple) orbital contributions are projected to the band. (**e**) The Brillouin zone and high symmetric path for *c*F. The bold lines indicate segments of the path, which is Γ-X-U | K-Γ-L-W-X.[4]